\documentclass[letterpaper,twocolumn,prl,aps,superscriptaddress,amsmath,amssymb,floatfix]{revtex4-1}
\usepackage{mathptmx}
\usepackage[latin9]{inputenc}
\usepackage{color}
\usepackage{amsmath}
\usepackage{amssymb}
\usepackage{graphicx}
\usepackage{esint}
\usepackage[english]{babel}
\usepackage[unicode=true,bookmarks=true,bookmarksnumbered=false,bookmarksopen=false,breaklinks=false,pdfborder={0 0 1},backref=false,colorlinks=true]{hyperref}
\hypersetup{linkcolor=magenta,urlcolor=blue,citecolor=blue,pdfstartview={FitH},hyperfootnotes=false}

\makeatletter

\usepackage{textcomp}
\usepackage{epstopdf}

\pdfpageheight\paperheight
\pdfpagewidth\paperwidth

\@ifundefined{textcolor}{}{%
 \definecolor{BLACK}{gray}{0}
 \definecolor{WHITE}{gray}{1}
 \definecolor{RED}{rgb}{1,0,0}
 \definecolor{GREEN}{rgb}{0,1,0}
 \definecolor{BLUE}{rgb}{0,0,1}
 \definecolor{CYAN}{cmyk}{1,0,0,0}
 \definecolor{MAGENTA}{cmyk}{0,1,0,0}
 \definecolor{YELLOW}{cmyk}{0,0,1,0}
}

\usepackage{xcolor}\usepackage{soul}
\definecolor{blue}{rgb}{0,0,1}
\definecolor{red}{rgb}{1,0,0}
\definecolor{green}{rgb}{0,1,0}

\usepackage{soul}

\makeatother

\begin{document}
\title{Circuit Quantum Acoustodynamics in a Scalable Phononic Integrated Circuit Architecture}

\author{Weiting~Wang}
\thanks{These four authors contributed equally to this work.}
\affiliation{Center for Quantum Information, Institute for Interdisciplinary Information Sciences, Tsinghua University, Beijing 100084, China}

\author{Lintao~Xiao}
\thanks{These four authors contributed equally to this work.}
\affiliation{Center for Quantum Information, Institute for Interdisciplinary Information Sciences, Tsinghua University, Beijing 100084, China}

\author{Bo~Zhang}
\thanks{These four authors contributed equally to this work.}
\affiliation{Center for Quantum Information, Institute for Interdisciplinary Information Sciences, Tsinghua University, Beijing 100084, China}

\author{Yu Zeng}
\thanks{These four authors contributed equally to this work.}
\affiliation{Laboratory of Quantum Information, University of Science and Technology of China, Hefei 230026, China.}
\affiliation{Anhui Province Key Laboratory of Quantum Network, University of Science and Technology of China, Hefei 230026, China}

\author{Ziyue Hua}
\affiliation{Center for Quantum Information, Institute for Interdisciplinary Information Sciences, Tsinghua University, Beijing 100084, China}

\author{Chuanlong Ma}
\affiliation{Center for Quantum Information, Institute for Interdisciplinary Information Sciences, Tsinghua University, Beijing 100084, China}

\author{Hongwei Huang}
\affiliation{Center for Quantum Information, Institute for Interdisciplinary Information Sciences, Tsinghua University, Beijing 100084, China}

\author{Yifang Xu}
\affiliation{Center for Quantum Information, Institute for Interdisciplinary Information Sciences, Tsinghua University, Beijing 100084, China}

\author{Jia-Qi~Wang}
\affiliation{Laboratory of Quantum Information, University of Science and Technology of China, Hefei 230026, China.}
\affiliation{Anhui Province Key Laboratory of Quantum Network, University of Science and Technology of China, Hefei 230026, China}

\author{Guangming~Xue}
\affiliation{Beijing Academy of Quantum Information Sciences, Beijing 100084, China}
\affiliation{Hefei National Laboratory, Hefei 230088, China}

\author{Haifeng~Yu}
\affiliation{Beijing Academy of Quantum Information Sciences, Beijing 100084, China}
\affiliation{Hefei National Laboratory, Hefei 230088, China}

\author{Xin-Biao~Xu}
\email{xbxuphys@ustc.edu.cn}
\affiliation{Laboratory of Quantum Information, University of Science and Technology of China, Hefei 230026, China.}
\affiliation{Anhui Province Key Laboratory of Quantum Network, University of Science and Technology of China, Hefei 230026, China}

\author{Chang-Ling~Zou}
\email{clzou321@ustc.edu.cn}
\affiliation{Laboratory of Quantum Information, University of Science and Technology of China, Hefei 230026, China.}
\affiliation{Anhui Province Key Laboratory of Quantum Network, University of Science and Technology of China, Hefei 230026, China}
\affiliation{Hefei National Laboratory, Hefei 230088, China}

\author{Luyan~Sun}
\email{luyansun@tsinghua.edu.cn}
\affiliation{Center for Quantum Information, Institute for Interdisciplinary Information
Sciences, Tsinghua University, Beijing 100084, China}
\affiliation{Hefei National Laboratory, Hefei 230088, China}

%\date{\today}
\begin{abstract}
Previous demonstrations of quantum acoustic systems have been limited to isolated devices, with limited capability to route phonons and interconnect multi-port acoustic elements for further extension. Here, we demonstrate a scalable architecture for circuit quantum acoustodynamics (cQAD) by integrating superconducting qubits with suspension-free phononic integrated circuits (PnICs). Coherent coupling between tunable transmon qubits and waveguide-integrated phononic cavities, including Fabry-Perot cavities via monolithic integration and microring cavities via flip-chip assembly, has been achieved, producing a pronounced enhancement of phonon emission with a Purcell factor of $\sim 19$. These devices represent elementary building blocks for scalable phononic circuits, establishing the foundation for phonon-based quantum information processors and the testbed for novel quantum acoustic phenomena.
\end{abstract}
\maketitle

\smallskip{}
\noindent \textbf{\large{}Introduction}{\large\par}
\noindent
Long-lived coherence, strong nonlinearity, and scalability have always been the core pursuits of quantum information platforms~\cite{CQED2021,Li2024Nature,Bland2025Nature}. While photons~\cite{obrien_optical_2007,zhong_quantum_2020,zhong_phase-programmable_2021,madsen_quantum_2022}, spins~\cite{wang_experimental_2017,hou_experimental_2019,zhang_observation_2021,kim_scalable_2025}, and superconducting circuits~\cite{sun_tracking_2014,arute_quantum_2019,grimm_stabilization_2020,google_ai_quantum_and_collaborators_hartree-fock_2020,zhao_realization_2022,acharya_suppressing_2023} each excel in specific metrics, satisfying all three simultaneously remains a challenge. Phonons, as quantized mechanical vibrations, offer alternative quantum information carriers that combines the advantage of short-wavelength~\cite{Felix2021PRA,Xu2022}, long coherence time~\cite{diamandi_optomechanical_2025,Bozkurt2025NatPhy}, and excellent compatibility with all existing quantum information carriers, thus holding promise as the core of a variety of hybrid quantum systems~\cite{Xiang2013,Barzanjeh2022}. In particular, acoustic waves at gigahertz frequencies possess micrometer-scale wavelengths~\cite{Felix2021PRA,Xu2022}, which are five orders of magnitude shorter than those of electromagnetic waves at the same energy. To advance phonon-based quantum platforms, great efforts have been devoted to develop coherent interfaces between phononic devices and individual qubits, including superconducting qubits~\cite{OConnell2010}, spin qubits in diamonds~\cite{Wang2020,Peng2025}, and semiconductor quantum dots~\cite{Mcneil2011}. Among these approaches, superconducting qubits coupled to acoustic resonators through piezoelectric~\cite{OConnell2010} or electromechanical coupling~\cite{Bozkurt2023NatPhy,Bozkurt2025NatPhy} have been explored to realize circuit quantum acoustodynamics (cQAD) platforms. These platforms demonstrate the capability to create, manipulate, and detect acoustic quantum states with unprecedented precision, and offer a promising path toward quantum information processors that leverage the unique advantages of phonons~\cite{Manenti2017NatCommun,Qiao2023,Yang2024Science,VonLupke2024}.

Despite these great advances, the realization of cQAD has been largely restricted to isolated acoustic devices, including bulk acoustic wave (BAW) resonators~\cite{Han2016,Yiwen2017Science,Yiwen2018Nature,Yang2024Science}, surface acoustic wave (SAW) devices~\cite{Martin2014Science,Sundaresan2015,Manenti2017,Satzinger2018Nature,Qiao2023,Ruan2024,Qiao2025}, and suspended membrane or beam-like structures~\cite{OConnell2010,Bozkurt2023NatPhy,Bozkurt2025NatPhy}, hindering further scalability of quantum acoustics. BAW resonators achieve the longest coherent times, reaching hundreds of microseconds to milliseconds, and show excellent potential for quantum memory applications. Nevertheless, their mode confinement across the substrate precludes the multi-port operation and direct phonon routing between devices. SAW devices have demonstrated phonon-mediated qubit entanglement and beamsplitter-type operations~\cite{Qiao2023,Qiao2025}, representing pioneering studies of multi-port and multi-device extensions, but the weak lateral confinement inherent to surface waves limits device density and prevents the realization of complex phononic circuits. Suspended structures offer strong mode confinement, but their scalability is severely limited by the challenges in mechanical fragility, fabrication complexity, and thermal management. Recently, suspension-free phononic integrated circuits (PnICs)~\cite{WeiFu2019NatCommun,Felix2021PRA,Xu2022} have emerged as a promising platform that can overcome these limitations. Analogous to photonic integrated circuits in optical systems, PnICs tightly confine gigahertz-frequency acoustic waves in micrometer-scale waveguides through acoustic index contrast, without requiring suspended structures. This architecture shows great potential for system-level phononic integration, but coherent coupling between superconducting qubits and PnIC-based resonators has not yet been demonstrated.

\begin{figure*}
\begin{centering}
\textcolor{red}{\includegraphics[width=1\linewidth]{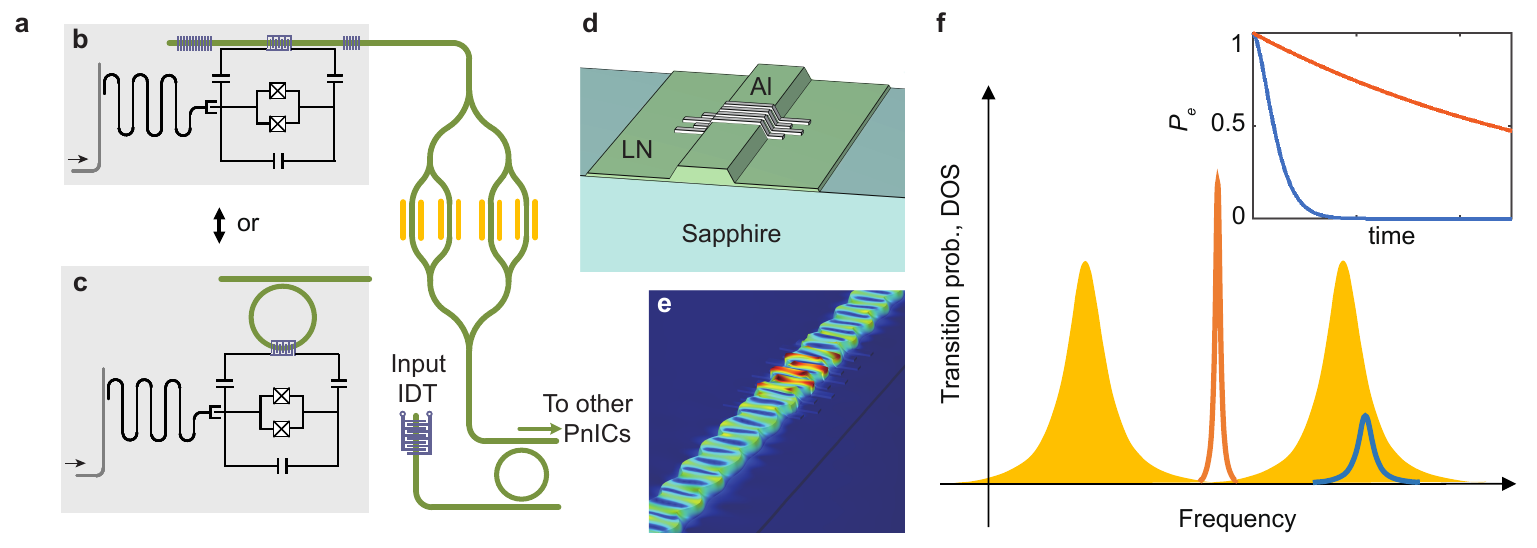}}
\par\end{centering}
\caption{\textbf{Principle of phononic integrated circuits (PnICs)-based circuit quantum acoustodynamics (cQAD) platform.} \textbf{a}, Schematic of the cQAD platform with a transmon coupled to a Fabry-Perot (FP) cavity (\textbf{b}) or a microring cavity (\textbf{c}). \textbf{d}, Cross-sectional schematic of the lithium niobate (LN) waveguide on a sapphire substrate with an interdigital transducer (IDT). \textbf{e}, Simulated mode profile showing tightly confined Love-like acoustic wave in the waveguide,  excited via the IDT. \textbf{f}, Principle of the Purcell effect in cQAD devices. Yellow regions denote the engineered phononic density of states (DOS) and solid lines represent the qubit transition frequencies. The transition linewidth broadens significantly when its frequency aligned with a cavity resonance (high DOS). Inset: The corresponding qubit decay dynamics is drastically accelerated by a phonon resonance.
}
\label{Fig1}
\end{figure*}

Here, we demonstrate the first scalable cQAD platform by integrating superconducting transmon qubits with suspension-free PnICs on lithium niobate-on-sapphire (LNOS) platform. We achieve coherent qubit-phonon coupling in two distinct cavity geometries: Fabry-Perot cavities realized through monolithic integration, and microring cavities implemented via flip-chip assembly. Both approaches yield robust phonon-qubit coupling, exhibit resonance-accelerated phonon emission of qubits through efficient piezoelectric coupling, and demonstrate Purcell factors up to $\sim19$. These results validate the PnIC-based cQAD platform by demonstrating the crucial coherent interface between the qubits and elementary waveguide-integrated devices, establishing the foundation for integration with other PnIC components and the realization of PnIC-based phononic quantum processors.

\smallskip{}
\noindent \textbf{\large{}Results}{\large\par}
%\noindent
\smallskip{}
\noindent {\textbf{Principle of PnICs-based cQAD devices} }
Figure~\ref{Fig1}a schematically illustrates the PnIC-based cQAD platform, consisting of phononic waveguides and associated devices. The waveguide structures are fabricated from an LNOS substrate. These suspension-free PnICs offer several advantages for scalability: superior thermal dissipation, mechanical robustness, and fabrication compatibility with standard chip processing. Furthermore, both single-crystal lithium niobate (LN) and sapphire exhibit excellent acoustic properties, potentially supporting high acoustic quality factors exceeding $10^7$. In contrast to the SAW devices where waves spread laterally, acoustic vibrations in PnICs are tightly confined within the LN wedge waveguides due to the lower acoustic velocity in LN ($\sim 3500 \,\mathrm{m/s}$) compared to sapphire ($\sim 5800\,\mathrm{m/s}$), similar to the mechanism of index contrast in photonic integrated circuits. The waveguide supports phonon modes dominated by out-of-plane (Rayleigh-like) and in-plane (Love-like) motion, resembling the two polarizations in photonic waveguides. For GHz-frequency acoustic waves, the corresponding wavelengths approach or even fall below one micron~\cite{WeiFu2019NatCommun,Felix2021PRA,Xu2022,Wang2022APL,Balram2024APL}. Recently, a versatile phononic device toolbox has been demonstrated, including multi-port directional couplers, beamsplitter, polarization converters, resonators, and filters, providing the essential elements for large-scale phonon-based information processing~\cite{xu_large-scale_2025}.

The interdigital transducer (IDT) serves as the key interface between electrical signals and acoustic waves via the piezoelectric effect. The RF voltage applied to the IDT electrodes can deform the LN, and the motion of LN can create electric fields across the electrode reciprocally. Beyond transduction, the periodic metallic electrode structure can also create a periodic modulation of the acoustic velocity through both piezoelectric stiffening and mass loading effects. This periodic perturbation induces distributed Bragg reflection (DBR) due to the bandgap effect, realizing a mirror for the traveling acoustic waves. The piezoelectric coupling also enables capacitive interaction between traveling acoustic waves and external circuit elements, such as superconducting resonators and transmon qubits, by converting mechanical displacement into voltage fluctuations on the IDT electrodes. Importantly, the anisotropic crystal structure of LN introduces direction-dependent acoustic properties, which must be carefully considered in device design, as it affects the IDT transduction efficiency, mirror reflectivity, and ultimately the qubit-phonon coupling strength.

\begin{figure*}[t]
\begin{centering}
\includegraphics[width=1\textwidth]{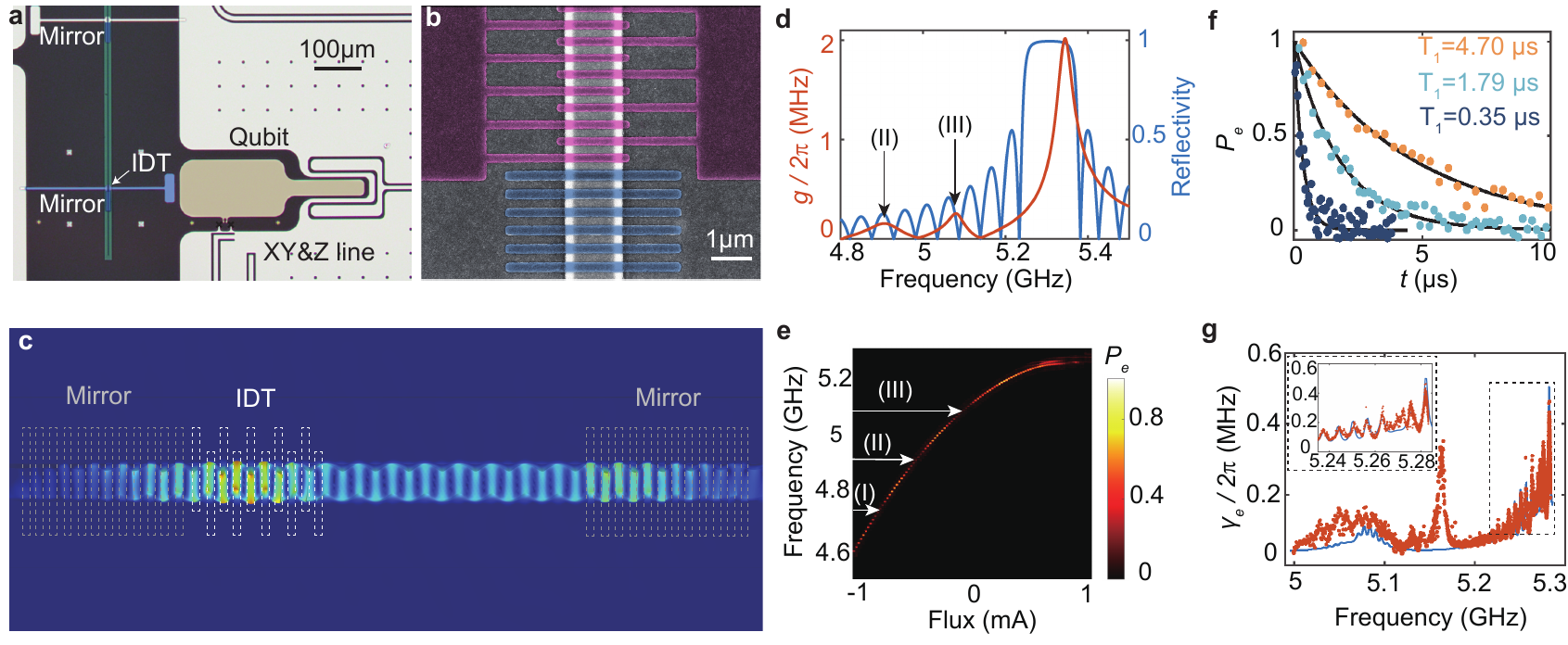}
\par\end{centering}
\caption{\textbf{Monolithically integrated cQAD device based on an FP phononic cavity.} \textbf{a}, False-color image of the device, which comprises a straight phononic waveguide (green), two mirrors (blue), and an IDT coupler (purple in \textbf{b}) for the transmon qubit. \textbf{b}, Magnified SEM picture of the IDT and a mirror. \textbf{c}, Simulated standing-wave mode profile confined by the two mirrors, using a re-scaled FP cavity structure for better visualization. \textbf{d}, Simulated qubit-cavity coupling strength and mirror reflectivity versus probe frequency. \textbf{e}, Qubit spectroscopy as a function of the flux bias. {Three frequencies marked I, II, and III correspond to near-zero measured populations. These frequencies are located within the lossy-cavity regime [regime (i), discussed in the main text]. The frequencies at positions II and III correspond to the two peaks of the coupling coefficient depicted in \textbf{d}.} \textbf{f}, Qubit decay dynamics at transition frequencies of $5.013\,\mathrm{GHz}$, $5.032\,\mathrm{GHz}$, and $5.281\,\mathrm{GHz}$, with fitted energy relaxation times of $4.70\,\mathrm{\mu s}$, $1.79\,\mathrm{\mu s}$, and $0.35\,\mathrm{\mu s}$, respectively. \textbf{g}, Measured qubit dissipation rate $\gamma_e/2\pi$ (red dots) and theoretical prediction (line) as functions of qubit frequency. Inset: expanded view near $5.27\,\mathrm{GHz}$ with a pronounced modulation due to phonon modes.
}
\label{Fig2}
\end{figure*}

We implement two distinct cavity geometries to demonstrate the versatility of the PnIC-cQAD platform, as illustrated in Fig.~\ref{Fig1}a. The first is a waveguide Fabry-Perot (FP) cavity (Fig.~\ref{Fig1}b), formed by placing two DBR mirrors along a straight waveguide segment with an IDT positioned between them. The mirrors provide frequency-selective reflection, confining acoustic energy within the cavity while the IDT enables electrical access to the cavity modes. The second geometry is a microring resonator (Fig.~\ref{Fig1}c), where a closed-loop waveguide supports whispering-gallery-like acoustic modes that couple evanescently to an adjacent bus waveguide. The ring geometry offers a particularly elegant approach to multi-port operation: acoustic elements can interact with the ring through evanescent coupling without physically interrupting the resonator structure, enabling both broadband confinement and flexible operation.

Both cavities support a series of resonances separated by the free spectral range (FSR) $\nu_\mathrm{FSR} = v_g/2L$, where $v_g$ is the phonon group velocity and $L$ is the cavity round-trip length. The coherent coupling between a transmon qubit and these modes is described by the Hamiltonian ($\hbar=1$)~\cite{Han2016}
\begin{equation}
H = \omega_q |e\rangle\langle e| + \sum_n \omega_n a_n^\dagger a_n + \sum_n  g_n (a_n^\dagger |g\rangle\langle e| + a_n |e\rangle\langle g|),
\label{Eq1}
\end{equation}
where $\omega_q$ is the qubit transition frequency, $\omega_n$
and $a_n$ ($a_n^\dagger$) are the frequency and annihilation (creation) operator of the
$n$-th cavity mode, respectively, and $g_n$ is the corresponding coupling strength. In the bad-cavity limit ($\kappa_n \gg g_n$), when the qubit is tuned near resonance with a cavity mode, it can emit phonons at a Purcell-enhanced rate $\gamma_n = 4g_n^2/\kappa_n$, where $\kappa_n$ is the cavity linewidth. Considering the contribution from all modes with detunings, the qubit's decay rate is
\begin{equation}
\gamma_{\text{e}}=\gamma_{\text{0}}+\sum_n \frac{4g_n^2\kappa_n }{4(\omega_q - \omega_n)^2 + \kappa_n^2},
\label{Eq2}
\end{equation}
where $ \gamma_0$ is the intrinsic dissipation rate of the qubit. The Purcell factor, $F_\mathrm{P} = \gamma_{\mathrm{e}}/\gamma_{0}$, quantifies the cavity-induced enhancement of spontaneous emission through the modification of phononic density of states (DOS). As shown in Fig.~\ref{Fig1}f, when the qubit frequency aligns with a cavity resonance, the locally enhanced DOS dramatically accelerates phonon emission. {This Purcell-enhanced emission is the hallmark signature of coherent qubit-phonon coupling in our cQAD platform.}

\smallskip{}
\noindent {\textbf{Monolithic-integration with an FP cavity} } We first demonstrate the cQAD platform through monolithic integration, where the transmon qubit and the phononic FP cavity are fabricated on a shared z-cut sapphire substrate. A ridge phononic waveguide is fabricated in the x-cut LN, featuring a thickness of $220\,\mathrm{nm}$, a width of $1\,\mathrm{\mu m}$, and 55 degree sidewalls, with a $50\,\mathrm{nm}$-thick slab layer surrounding the waveguide, as shown in Fig.~\ref{Fig1}d. Outside of the slab layer, all the LN is completely etched away, and the superconducting (aluminum) electrodes and the transmon qubit are then fabricated on the exposed sapphire.

The devices and detailed structures are depicted in Figs.~\ref{Fig2}a and \ref{Fig2}b. The phononic FP cavity is constructed from a straight waveguide, with two DBR mirrors separated by $L = 300\,\mathrm{\mu m}$. As detailed in Fig.~\ref{Fig2}b, each DBR mirror comprises 100 Al strips with a period of 430\,nm and a duty cycle of 0.5, and an IDT with 20 finger pairs with a 782\,nm period is placed between the mirrors. One IDT electrode connects directly to ground, while the other extends to a floating metallic pad for capacitive coupling with the transmon qubit. The qubit state is detected through a superconducting readout resonator that dispersively couples to the qubit. To illustrate how the phononic energy is localized within the LN waveguide FP cavity, we simulate an acoustic eigenmode in a rescaled structure with much shorter length and fewer electrode fingers. The standing wave pattern formed between the two DBR mirrors clearly demonstrates the capability to effectively reflect acoustic waves and form tightly confined mode, as shown in Fig.~\ref{Fig2}c. Since both the IDT and mirrors have a periodic structure, their responses show frequency-dependence. Figure~\ref{Fig2}d presents the numerically calculated mirror reflectivity and qubit-cavity coupling strength $g$. The value of $g$ is obtained from simulation with an overall scaling factor extracted from experimental fitting (see below). Both quantities reach their optimal values near $5.34\,\mathrm{GHz}$, confirming well-aligned operating frequencies.

The cQAD system is experimentally characterized by measuring the qubit excitation population ($P_e$) spectrum under a pulsed microwave drive while tuning the qubit transition frequency via its flux bias, as shown in Fig.~\ref{Fig2}e. As the qubit frequency sweeps from $4.60\,\mathrm{GHz}$ to $5.26\,\mathrm{GHz}$, the spectrum reveals pronounced periodic modulations in both the transition linewidth and readout contrast, indicating the coherent coupling between the qubit and the phononic modes. This periodic pattern is attributed to the frequency-dependent sinc-function response of the IDT, in excellent agreement with the theoretical prediction in Fig.~\ref{Fig2}d. Time-resolved measurements provide direct evidence of the Purcell-enhanced phonon emission. Figure~\ref{Fig2}f compares the decay dynamics of the qubit at frequencies of $5.013\,\mathrm{GHz}$, {$5.032\,\mathrm{GHz}$} and $5.281\,\mathrm{GHz}$, representing the significant modification of the qubit's spontaneous decay by structured {phononic vacuum}. Despite a small frequency difference of only 6\,MHz, the decay rates differ dramatically, manifesting the strong Purcell effect by our engineered phononic DOS in our PnICs platform. The accelerated decay corresponds to a cavity-enhanced phonon emission rate up to {$2.67\, \mathrm{MHz}$}, implying that by driving the qubit to the excited state, we have a probability of {$92.7\%$} for generating single phonons into the PnICs, with a short single phonon pulse duration of {$375\,\mathrm{ns}$} (or a pulse length of {$1350\,\mathrm{\mu m}$}) in the waveguide.

The Purcell effect arising from the structured phonon bath is systematically investigated by finely scanning the qubit frequency, as shown in Fig.~\ref{Fig2}g. Using the theoretically predicted mirror reflectivity and coupling strengths (Fig.~\ref{Fig2}d), along with the multimode cQAD model in Eq.~(\ref{Eq2}), we calculate the qubit decay rate $\gamma_\mathrm{e}$ with fitted $\gamma_0$ for comparison with experimental data. The fitted bare qubit decay rate $ \gamma_0 $ corresponds to a $T_1$ time of $4.8\,\mathrm{\mu s}$. This value is consistent with the qubit's lifetime when the coupling coefficient is at its minimum (at $\omega_q=5.00$~GHz, where $g\approx0$, the measured qubit lifetime $T_1=4.7\,\mathrm{\mu s}$).

Across the wide spectral range of 300\,MHz, the cQAD dynamics can be divided into three distinct regimes: (i) Lossy cavity regime ($\omega_q<5.12\,\mathrm{GHz}$). Although enhanced decay is present due to non-vanishing IDT coupling, the mirror reflectivity is too low to support well-confined phononic cavity modes. In this regime, the system resembles waveguide QAD dynamics, where the qubit couples to a quasi-continuum of propagating phonon modes, yielding a maximum Purcell factor of {$\sim 5.0$}. (ii) Anomalous regime ($\omega_q/2\pi \approx 5.16\,\mathrm{GHz}$). A pronounced enhancement in decay rate emerges even though the coupling $g$ is predicted to be negligible at this frequency. This feature cannot be explained by the standard cQAD model, suggesting the presence of an additional qubit-phonon coupling mechanism in our system that warrants further investigation. (iii) cQAD regime  ($\omega_q/2\pi > 5.18\,\mathrm{GHz}$). In this regime, both $g$ and mirror reflectivity approach their optimal values. We observe a pronounced periodic modulation of the decay rates that matches the cavity's FSR, in excellent agreement with our theoretical model. The fitted FSR of the cavity modes is 6.4~MHz, corresponding to group velocity of $v_\mathrm{g}=3840\,\mathrm{m/s}$ which is consistent with the calculated value of 6~MHz based on numerically predicted $v_\mathrm{g}=3600\,\mathrm{m/s}$ for the Love mode. The largest Purcell factor of $\sim14.0$ is achieved in this regime, and the fitted quality factor of the cavity mode is $2.2\times10^3$. Our simulations predict a maximum $g/2\pi\approx2.1\,\mathrm{MHz}$ at the IDT's optimal working frequency around $5.35\,\mathrm{GHz}$, indicating that the strong coupling regime, where $g$ exceeds the decay rates of both the qubit and the phonon mode, is achievable. However, the limited tuning range of the current qubit design (Fig.~\ref{Fig2}f) prevents exploration of this regime, which motivates future device optimization.

\begin{figure*}[t]
\begin{centering}
\textcolor{red}{\includegraphics[width=1\linewidth]{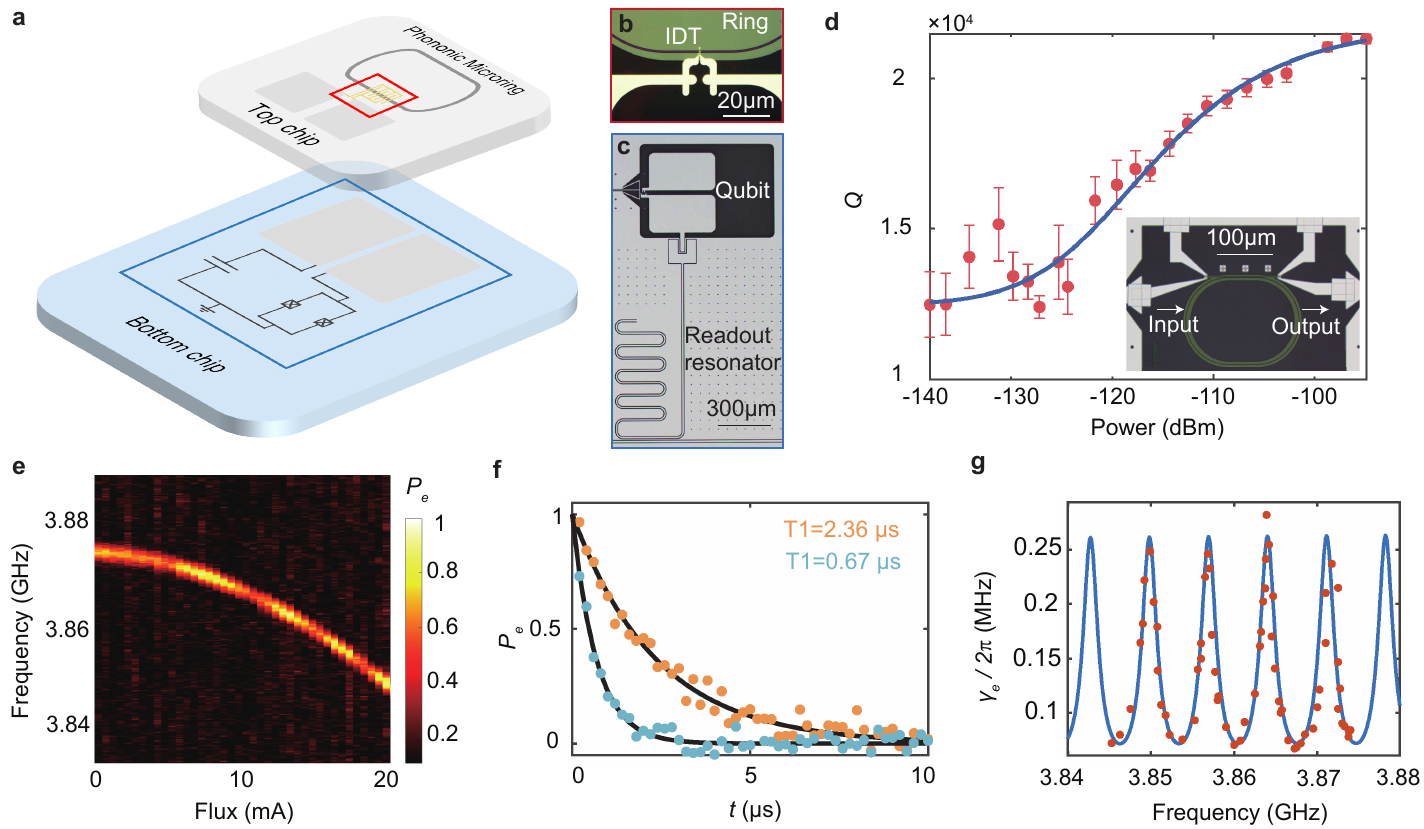}}
\par\end{centering}
\caption{\textbf{cQAD device based on a microring cavity.} \textbf{a}, Schematic of the flip-chip assembly. The phononic microring cavity is fabricated on a top chip of 11~mm by 11~mm, while the superconducting qubit along with its readout and control circuitry are fabricated on a bottom chip of 15~mm by 15~mm. \textbf{b}, Schematic of qubit-ring cavity coupling. An IDT is placed on the ring cavity, with its two electrodes connected to two large pads. These pads form a coupling capacitance with the pads of the floating transmon, enabling the interaction between the qubit and the cavity mode. \textbf{c}, Optical micrograph of the bottom chip. \textbf{d}, Quality factor of the phononic microring cavity as a function of the input power to the chip. \textbf{e}, Qubit spectroscopy as a function of flux bias. \textbf{f}, Qubit decay dynamics at two distinct frequencies: 3.867~GHz (resonant, blue points) with $T_1=0.67\,\mathrm{\mu s}$ and 3.872~GHz with $T_1=2.36\,\mathrm{\mu s}$ (off-resonant, orange points). \textbf{g}, Effective qubit dissipation rate as a function of transition frequency. Experimental data are shown as dots; the solid line represents the theoretical fit.
}
\label{Fig3}
\end{figure*}

\smallskip{}
\noindent {\textbf{Flip-chip assembly with a microring} } While the monolithic integration approach ensures precise alignment and direct coupling between devices, the superconducting qubit's coherence may be compromised by residual LN clusters remaining on the sapphire surface after fabrication. To circumvent this issue and demonstrate the versatility of our platform, we develop a complementary flip-chip approach, which enables high-yield integration between PnICs and other mature quantum circuit technologies for further scalability.

Figure~\ref{Fig3}a illustrates our flip-chip device, which combines a PnIC chip hosting a microring cavity and a separated superconducting qubit chip. The two chips are assembled face-to-face with a precisely controlled vertical gap of about $10\,\mathrm{\mu m}$, enabling capacitive coupling while maintaining physical isolation between the two chips. The microring has a racetrack shape, with only a single pair of IDT fingers placed on the straight section. The IDT electrodes are connected to two metallic pads for capacitive coupling with the corresponding metallic pads across the vacuum gap, thereby interconnecting the qubit on the other chip.

Unlike the FP cavity, which relies on DBR mirrors and is restricted by limited working bandwidth, the microring resonators have a confinement mechanism that is less sensitive to the phonon wavelength and temperature. This enables high-quality traveling-wave whispering-gallery-like acoustic modes over a large bandwidth. Furthermore, the microring can be coupled to external bus phononic waveguides for further extension~\cite{Xu2022}. Before investigating cQAD dynamics, we first characterize the phonon modes at the single-quanta level in an isolated microring (inset of Fig.~\ref{Fig3}d) through the bus waveguide {at about $5.4\,\mathrm{GHz}$ with varying power}. As shown in Fig.~\ref{Fig3}d, the resonance quality factor exceeds $2.1\times10^4$ at high probe powers, but degrades and saturates to $1.2\times10^4$ at the single-quanta level, indicating the loss due to two-level systems in the material~\cite{Wollack2021,Gruenke-Freudenstein2025}. We also measure the lifetimes of three isolated qubits {fabricated without flip-chip integration}, obtaining relaxation times of {$34.5\,\mathrm{\mu s}$}, {$46.9\,\mathrm{\mu s}$} and {$64.5\,\mathrm{\mu s}$}, respectively.

The corresponding cQAD dynamics of the combined chips are investigated in Figs.~\ref{Fig3}e and \ref{Fig3}f. Distinct from the FP cavity case, the single-pair IDT exhibits a broad working bandwidth, and the modulation in the excitation spectra in Fig.~\ref{Fig3}e is mainly attributed to the equally-spaced phonon modes in the microring. Comparing the typical decay dynamics at $3.867\,\mathrm{GHz}$ (on-resonance) and $3.872\,\mathrm{GHz}$ (off-resonance), as shown in Fig.~\ref{Fig3}f, we extract the qubit relaxation lifetimes of $T_1=0.67\,\mathrm{\mu s}$ and $2.36\,\mathrm{\mu s}$, respectively, providing direct evidence of the Purcell-enhanced phonon emission into the microring cavity.

Figure~\ref{Fig3}g displays the systematic frequency dependence of qubit's dissipation rate $\gamma_\mathrm{e}$. The pronounced periodic modulation pattern reflects the cavity mode structure, which exhibits equal frequency spacing, as well as comparable coupling strengths and quality factors across modes. From the measurements, we extract $\nu_\mathrm{FSR}=7.1\,\mathrm{MHz}$, corresponding to a mean group velocity of $4050\,\mathrm{m/s}$ that is consistent with the expected value given the anisotropic effects around the ring circumference. By fitting the frequency-dependent dissipation rate to the multimode cQAD model, shown as the blue curve in Fig.~\ref{Fig3}g, we extract the key system parameters: a uniform cavity quality factor $Q=1.7\times10^3$, a cross-chip qubit-phonon mode coupling strength $g/2\pi=0.36\,\mathrm{MHz}$, and an intrinsic qubit decay rate $\gamma_0/2\pi={0.0147}\,\mathrm{MHz}$ corresponding to {$T_1 = 10.8\,\mu s$}. The on-resonance coupling yields Purcell factors up to {19.2}, resulting in a single-phonon generation probability of {$94.7\%$} through the qubit.

Compared with the isolated devices, we find that both the values of microring $Q$ and the qubit $T_1$ are reduced in the bounded flip-chip assembly, though both metrics still exceed those of the monolithically integrated devices in Fig.~\ref{Fig2}. These results confirm that the residual LN on the substrate or other aspects of LN processing introduce defects and degrade the qubit coherence. By comparing the devices with and without IDTs, we identify an additional loss mechanism arising from the non-ideality of the IDT coupler. The IDTs simultaneously couple to target guided modes in the waveguide and to a continuum of leaky surface and bulk phonon modes in the slab and substrate. This continuum exhibits a broadband response, so non-ideal coupling introduces extra frequency-insensitive decay channels for both cavity modes and qubits, contributing to the degraded phonon $Q$ and qubit $T_1$ observed in the coupled devices. We emphasize that the intrinsic losses derived from our cQAD model already account for these non-ideality-induced decay rates. The reported Purcell factor therefore quantifies the enhancement of spontaneous emission specifically into the target guided modes, representing the relevant figure of merit for potential phononic applications.

\smallskip{}
\noindent \textbf{\large{}Discussion}{\large\par}
\noindent Our work establishes a scalable architecture for phononic quantum information processing. Both monolithic integration and flip-chip assembly achieve high-cooperativity cQAD effects, demonstrating the versatility and robustness of our quantum PnIC platform. Combined with the existing PnIC component toolbox~\cite{Xu2025NE,xu_large-scale_2025}, this platform enables the construction of complex multi-qubit phononic networks rather than isolated devices. Beyond on-chip scalability, this material platform is inherently compatible with coherent phonon-photon frequency conversion, enabling the realization of ``Zhengfu" architecture for hybrid superconducting-phononic-photonic chips~\cite{Xu2022b}. By introducing optical interfaces, this architecture could facilitate quantum interconnections between distant chips even in separate dilution refrigerators~\cite{Yang2025,Zou2025}, providing a pathway toward distributed quantum computing.

Our system approaches the strong coupling regime, currently limited primarily by phonon dissipation arising from fabrication imperfections. Given that single-crystal LN and sapphire substrates have demonstrated acoustic quality factors approaching $10^7$~\cite{Yiwen2018Nature,Wollack2021,Gruenke-Freudenstein2025}, substantial improvements are achievable through optimized fabrication processing. Furthermore, the coupling strength $g$ can be enhanced by increasing the number of IDT finger pairs and optimizing qubit design. Since our devices already exhibit multimode cQAD dynamics~\cite{VonLupke2024}, pushing $g$ beyond the FSR would access the intriguing superstrong coupling regime~\cite{Sundaresan2015,Moores2018}, where a single qubit simultaneously interacts with multiple phonon modes. Notably, our measurements reveal an anomalous spontaneous emission peak that cannot be explained by FP cavity modes alone, suggesting additional phonon coupling channels that merit further investigation. Our demonstration therefore not only provides a scalable cQAD platform for quantum information processing~\cite{Hann2019,Qiao2023}, but also offers a rich testbed for exploring novel quantum acoustic phenomena~\cite{Zou2025NP,Odeh2025,Kuruma2025} and testing fundamental physics~\cite{Schrinski2023,Linehan2025}.

\smallskip{}

\noindent \textbf{\large{}Acknowledgment}{\large\par}

\noindent This work was funded by the Innovation Program for Quantum Science and Technology (Grant Nos.~2024ZD0301500 and 2021ZD0300200) and the National Natural Science Foundation of China (Grant Nos.~92265210, 123B2068, 12104441, 12474498, 92165209, 12293053, 12374361, 92365301, and 92565301). The numerical calculations in this paper were performed on the supercomputing system in the Supercomputing Center of USTC, and this work was partially carried out at the USTC Center for Micro and Nanoscale Research and Fabrication.

\end{document}